\documentclass{jnmp}

\usepackage{amsmath}
\usepackage{latexsym}

\JNMPnumberwithin{equation}{section}

\newtheorem{theorem}{Theorem}[section]
\newtheorem{lemma}{Lemma}[section]
\newtheorem{proposition}{Proposition}[section]

\theoremstyle{definition}

\newtheorem{remark}{Remark}[section]
\begin{document}

%  Headings
%
\renewcommand{\evenhead}{Yu B Suris}
\renewcommand{\oddhead}{Integrable Discretizations
of Some Cases of the Rigid Body Dynamics}

%  Titlepage
%
\thispagestyle{empty}

\FirstPageHead{8}{4}{2001}{\pageref{suris-firstpage}--\pageref{suris-lastpage}}{Article}
%  Parameters: Volume, number, year, page range, paper type
%  'Article' could be changed to 'Letter' or 'Review Article'

\copyrightnote{2001}{Yu B Suris}

\Name{Integrable Discretizations of Some Cases\\
of the Rigid Body Dynamics}
\label{suris-firstpage}

\Author{Yuri B SURIS}

\Address{Fachbereich Mathematik, Technische Universit\"at Berlin, \\
Str. 17 Juni 136, 10623 Berlin, Germany\\
E-mail: suris@sfb288.math.tu-berlin.de}

\Date{Received April 12, 2001; Accepted June 4, 2001}

\begin{abstract}
\noindent
A heavy top with a fixed point and a rigid body in an ideal fluid are
important examples of Hamiltonian systems on a dual to the semidirect product
Lie algebra ${\rm e}(n)={\rm so}(n)\ltimes{\mathbb R}^n$. We give a Lagrangian
derivation of the corresponding equations of motion, and introduce discrete
time analogs of two integrable cases of these systems: the Lagrange top and
the Clebsch case, respectively. The construction of discretizations is
based on the discrete time Lagrangian mechanics on Lie groups, accompanied
by the discrete time Lagrangian reduction. The resulting explicit maps
on ${\rm e^*}(n)$ are Poisson with respect to the Lie--Poisson
bracket, and are also completely integrable. Lax representations of these maps
are also found.
\end{abstract}

\section{Introduction}
The rigid body dynamics are rich with problems interesting from the
mathematical point of view, in particular, with integrable problems.
Certainly, the most famous ones are the three integrable cases of the
rotation of a heavy rigid body around a fixed point in a homogeneous
gravity field, named after Euler, Lagrange, and Kovalevskaya. Another
problem with a number of integrable cases is the Kirchhoff's one, dealing
with the motion of a rigid body in an ideal fluid. Here integrable cases
carry the names of Kirchhoff, Clebsch, Steklov and Lyapunov. All these
classical cases were discovered in the 18th and the 19th century. However,
the list of intergrable problems of the rigid body dynamics is by far not
exhausted by these ones. For example, the integrablity of equations describing
the rotation
of a rigid body around its fixed center of mass in an arbitrary quadratic
potential is a much more recent observation due to Reyman \cite{Re} and
Bogoyavlensky~\cite{Bo2}. (However, some particular case of this result was
given already by Brun~\cite{Br}; the equations of motion in this case
are identical with those describing the integrable case of the
motion of a rigid body in an ideal fluid, due to Clebsch~\cite{C}).

It was realized already in the 19th century that some of these problems
admit multi-dimensional generalizations. According to bibliographical
remarks in~\cite{Bo2} and in~\cite{FK}, the multi-dimensional
generalization of the Euler top is due to Frahm~\cite{Fr} and Schottky
\cite{Sch}. A~modern period of the interest in this system started with the
general theory of geodesic flows on Lie groups, due to Arnold~\cite{Ar1},
as well as with the paper by Manakov~\cite{M}, who found the $n$-dimensional
Euler top as a reduction of the so called $n$ wave equations, along with a
spectral parameter dependent Lax representation. This was followed by a cycle
of papers by Mishchenko and Fomenko and their school, generalizing the
construction of a rigid body for a wide class of Lie algebras, see a review
of this work in~\cite{TF}.

As for the multi-dimensional
analogs of the Lagrange top, there exist two different versions. The first one, living on
${\rm so}(n)\ltimes{\mathbb R}^n$, was proposed in \cite{Be,Re}, see
also~\cite{RS2}. The second one, living on ${\rm so}(n)\ltimes{\rm so}(n)$,
was introduced in~\cite{Ra}.
The multi-dimensional generalization of the Kovalevskaya top is due to
\cite{RS1}, see also \cite{RS2,BRS}. For the Clebsch problem,
the multi-dimensional generalization was found in \cite{P},
for the Steklov--Lyapunov case a multi-dimensional generalization, living
on ${\rm so}(n)\ltimes{\rm so}(n)$ and not on ${\rm e}(n)={\rm so}(n)\ltimes
{\mathbb R}^n$, was found in~\cite{BF}.

In the present paper, we find {\it integrable discretizations} for the
multi-dimensional Lagrange top and for a certain flow of the integrable
hierarchy of the multi-dimensional Clebsch case. Our construction
is based on the discrete time Lagrangian mechanics, as pioneered by
Moser and Veselov \cite{MV}, and further developed in \cite{BS1,BS2,S1}.
In order to give the reader a flavour of our
integrable discretizations, we present here the corresponding formulas.
In these formulas $(M,P)\in{\rm e}^*(n)$, i.e. $M\in{\rm so}(n)$,
$P\in\mathbb R^n$.
The equations of motion of the totally symmetric Lagrange top in the moving
frame (which are also equivalent to the equations of motion of an arbitrary
Lagrange top in the rest frame) and their integrable discretization are
($A\in\mathbb R^n$ is a constant vector):
\[
\left\{\begin{array}{l}
\dot{M}= A\wedge P,\vspace{2mm}\\
\dot{P}=-MP,
\end{array}\right. \qquad
\left\{\begin{array}{l}
M_{k+1}=M_k+\epsilon A\wedge P_k,\vspace{2mm}\\
\displaystyle P_{k+1}=\frac{I-(\epsilon/2) M_{k+1}}{I+(\epsilon/2)M_{k+1}}\,P_k.
\end{array}\right.
\]
Similarly, the equations of the Clebsch case of the rigid body motion in an
ideal fluid and their integrable discretization read ($B={\rm diag}\,
(b_1,\ldots,b_n)$ is a constant matrix):
\[
\left\{\begin{array}{l} \dot{M}=P\wedge (BP),\vspace{2mm}\\
\dot{P}=-MP,\end{array}\right.
\qquad
\left\{\begin{array}{l}
\displaystyle M_{k+1}=M_k+\epsilon \frac{P_k\wedge(BP_k)}
{1+(\epsilon^2/4)\langle P_k,BP_k\rangle},\vspace{2mm}\\
\displaystyle P_{k+1}=\frac{I-(\epsilon/2) M_{k+1}}{I+(\epsilon/2)M_{k+1}}\,P_k.
\end{array}\right.
\]

We recall the general theory of the Lagrangian reduction in the continuous time
and in the discrete time contexts, respectively, in Section~2 and~3. Further,
we give in Section~4 a Lagrangian derivation of the equations of motion of a
multi-dimensional heavy top, and in Section~5 we discuss the Lagrangian
theory of the multi-dimensional Lagrange top, both in the moving frame and
in the rest frame. Section~6 is devoted to the construction of an integrable
Lagrangian discretization of the multi-dimensional Lagrange top, again in
both formulations (in the moving frame and in the rest frame). Finally, in
Section~7 we discuss the Lagrangian formulation of the problem of the rigid
body motion in an ideal fluid, its integrable case due to Clebsch, and
construct an integrable Lagrangian discretization of this problem.
Conclusions are contained in Section~8.

\newpage

\section{Lagrangian mechanics and Lagrangian reduction on \mbox{\mathversion{bold}$TG$}}

Recall that a continuous time Lagrangian system is defined by a smooth function
${\mathbf L}(g,\dot{g}):TG\mapsto{\mathbb R}$ on the tangent bundle of a smooth
manifold $G$. The function ${\mathbf L}$ is called the {\it Lagrange function}. We
will be dealing here only with the case when $G$ carries an additional
structure of a {\it Lie group}, with the Lie algebra ${\mathfrak g}$.
For an arbitrary function $g(t):[t_0,t_1]\mapsto G$ one can consider
the {\it action functional}
\begin{equation*}%\label{action}
{\mathbf S}=\int_{t_0}^{t_1}{\mathbf L}(g(t),\dot{g}(t))dt.
\end{equation*}
A standard argument shows that the functions $g(t)$ yielding extrema of this
functional (in the class of variations preserving $g(t_0)$ and $g(t_1)$),
satisfy with necessity the {\it Euler--Lagrange equations}. In local
coordinates $\{g^i\}$ on $G$ they read:
\begin{equation}\label{EL gen}
\frac{d}{dt}\left(\frac{\partial{\mathbf L}}{\partial\dot{g}^i}\right)=
\frac{\partial{\mathbf L}}{\partial g^i}.
\end{equation}
The action functional ${\mathbf S}$ is independent on the choice of local coordinates,
and thus the Euler--Lagrange equations are actually coordinate independent as
well. For a coordinate-free description in the language of differential
geometry, see \cite{Ar1,MR}.

Introducing the quantities
\begin{equation*}%\label{Pi}
\Pi=\nabla_{\dot{g}}{\mathbf L}\in T_g^* G,
\end{equation*}
one defines the {\it Legendre transformation}:
\begin{equation*}%\label{Legendre gen}
(g,\dot{g})\in TG\;\mapsto\; (g,\Pi)\in T^*G.
\end{equation*}
If it is invertible, i.e. if $\dot{g}$ can be expressed through $(g,\Pi)$,
then the Legendre transformation of the Euler--Lagrange equations
(\ref{EL gen}) yield  a {\it Hamiltonian system} on $T^*G$ with respect to the
standard symplectic structure on $T^*G$ and with the Hamilton function
\begin{equation}\label{Ham gen}
H(g,\Pi)=\langle \Pi,\dot{g}\rangle-{\mathbf L}(g,\dot{g}),
\end{equation}
(where, of course, $\dot{g}$ has to be expressed through $(g,\Pi)$).

When working with the tangent bundle of a Lie group, it is convenient
to trivialize it, translating all vectors to the group unit by left or right
multiplication. We consider first the left trivialization:
\begin{equation}\label{left triv}
(g,\dot{g})\in TG\;\mapsto\;(g,\Omega)\in G\times{\mathfrak g},
\end{equation}
where
\begin{equation*}%\label{Omega}
\Omega=L_{g^{-1}*}\dot{g} \quad\Leftrightarrow\quad \dot{g}=L_{g*}\Omega.
\end{equation*}
Denote the Lagrange function pushed through the above map, by
${\mathbf L}^{(l)}(g,\Omega): G\times{\mathfrak g}\mapsto{\mathbb R}$, so that
\begin{equation*}%\label{Lagr left}
{\mathbf L}^{(l)}(g,\Omega)={\mathbf L}(g,\dot{g}),
\end{equation*}

The trivialization (\ref{left triv}) of the tangent bundle $TG$ induces
the following trivialization of the cotangent bundle $T^*G$:
\begin{equation*}%\label{left triv *}
(g,\Pi)\in T^*G\;\mapsto\;(g,M)\in G\times{\mathfrak g}^* ,
\end{equation*}
where
\begin{equation*}%\label{M}
M=L_g^*\Pi \quad\Leftrightarrow\quad \Pi=L_{g^{-1}}^* M.
\end{equation*}

We now consider the Lagrangian reduction procedure, in the case when the
Lagrange function is symmetric with respect to the left action of a certain
subgroup of $G$, namely an isotropy subgroup of some element in the
representation space of $G$. So, the next ingredient of our construction is a
representation $\Phi: G\times V\mapsto V$ of a Lie group $G$ in a linear
space $V$; we denote it by
\[
\Phi(g)\cdot v \quad{\rm for} \quad g\in G,\;\; v\in V.
\]
We denote also by $\phi$ the corresponding representation of the Lie algebra
${\mathfrak g}$ in $V$:
\begin{equation*}
\phi(\xi)\cdot v=\left.\frac{d}{d\epsilon}\Big(\Phi(e^{\epsilon\xi})\cdot
v\Big)\right|_{\epsilon=0} \qquad {\rm for} \quad \xi\in{\mathfrak g}, \;\; v\in V.
\end{equation*}
The map $\phi^*:{\mathfrak g}\times V^*\mapsto V^*$  defined by
\begin{equation*}%\label{phi star}
\langle \phi^*(\xi)\cdot y,v\rangle=\langle y,\phi(\xi)\cdot v\rangle\qquad
\forall \; v\in V,\; y\in V^*,\;\xi\in{\mathfrak g},
\end{equation*}
is an anti-representation of the Lie algebra ${\mathfrak g}$ in $V^*$.
We shall use also the bilinear operation $\diamond: V^*\times V\mapsto{\mathfrak g}^*$
defined as follows: let $v\in V$, $y\in V^*$, then
\begin{equation*}%\label{diamond op}
\langle y\diamond v, \xi\rangle=-\langle y,\phi(\xi)\cdot v\rangle\qquad
\forall \; \xi\in{\mathfrak g}.
\end{equation*}
(Notice that the pairings on the left-hand side and on the right-hand side
of the latter equation are defined on different spaces. The operation $\diamond$
can be found in~\cite{K}, where it plays the key role in the theory of the so
called Clebsch representations; our notation follows \cite{HMR,CHMR}.)

Fix an element $p\in V$, and consider the isotropy subgroup $G^{[p]}$ of $p$,
i.e.
\begin{equation*}%\label{G[a]}
G^{[p]}=\{h:\Phi(h)\cdot p=p\}\subset G.
\end{equation*}
Suppose that the Lagrange function ${\mathbf L}(g,\dot{g})$ is invariant under
the action of $G^{[p]}$ on $TG$ induced by left translations on $G$:
\begin{equation*}%\label{left action}
{\mathbf L}(hg,L_{h*}\dot{g})={\mathbf L}(g,\dot{g}), \qquad h\in G^{[p]}.
\end{equation*}
The corresponding invariance property of ${\mathbf L}^{(l)}(g,\Omega)$ is expressed as:
\begin{equation*}%\label{left action for red}
{\mathbf L}^{(l)}(hg,\Omega)={\mathbf L}^{(l)}(g,\Omega), \qquad h\in G^{[p]}.
\end{equation*}
We want to reduce the Euler--Lagrange equations with respect to this left
action. As a section $(G\times{\mathfrak g})/G^{[p]}$ we choose the set
${\mathfrak g}\times O_{p}$, where $O_{p}$ is the orbit of $p$ under the action~$\Phi$:
\begin{equation*}%\label{orbit}
O_p=\{\Phi(g)\cdot p,\;g\in G\}\subset V.
\end{equation*}
The reduction map is
\begin{equation*}
(g,\Omega)\in G\times{\mathfrak g}\;\mapsto\; (\Omega,P)\in {\mathfrak g}\times O_p, \qquad
{\rm where}  \quad P=\Phi(g^{-1})\cdot p,
\end{equation*}
so that the reduced Lagrange function ${\mathcal L}^{(l)}:{\mathfrak g}\times O_p
\mapsto{\mathbb R}$ is defined as
\begin{equation*}%\label{left red Lagr}
{\mathcal L}^{(l)}(\Omega,P)={\mathbf L}^{(l)}(g,\Omega),\qquad{\rm where}\quad
 P=\Phi(g^{-1})\cdot p.
\end{equation*}
The reduced Lagrangian ${\mathcal L}^{(l)}(\Omega,P)$ is well defined, because from
\[
P=\Phi\left(g_1^{-1}\right)\cdot p=\Phi\left(g_2^{-1}\right)\cdot p
\]
there follows $\Phi\left(g_2g_1^{-1}\right)\cdot p=p$, so that $g_2g_1^{-1}\in G^{[p]}$,
and ${\mathbf L}^{(l)}(g_1,\Omega)={\mathbf L}^{(l)}(g_2,\Omega)$.

\begin{theorem}\label{continuous reduction left} {\rm \cite{HMR,CHMR}}

{\rm a)} Under the left trivialization $(g,\dot{g})\mapsto(g,\Omega)$ and the
subsequent reduction $(g,\Omega)\mapsto(\Omega,P)$, the Euler--Lagrange
equations (\ref{EL gen}) become the following {\bf left Euler--Poincar\'e
equations}:
\begin{equation}\label{EL left Ham red}
\left\{\begin{array}{l}
\dot{M}={\rm ad}^*\,\Omega\cdot M+\nabla_P{\mathcal L}^{(l)}\diamond P,\vspace{1mm}\\
\dot{P}=-\phi(\Omega)\cdot P,\end{array}\right.
\end{equation}
where
\begin{equation*}%\label{M thru L red}
M=L_g^*\Pi=\nabla_{\Omega}{\mathcal L}^{(l)}.
\end{equation*}

{\rm b)} If the ``Legendre transformation''
\begin{equation*}%\label{Legendre left red}
(\Omega,P)\in {\mathfrak g}\times O_p\mapsto (M,P)\in {\mathfrak g}^*\times O_p,
\end{equation*}
is invertible, then (\ref{EL left Ham red}) is a Hamiltonian system
on ${\mathfrak g}^*\times O_p$ with the Hamilton function
\[
H(M,P)=\langle M,\Omega\rangle-{\mathcal L}^{(l)}(\Omega,P),
\]
with respect to the Poisson bracket given by
\begin{gather}
\{F_1,F_2\}=\langle M,[\nabla_M F_1,\nabla_M F_2]\,\rangle\nonumber\\
\phantom{\{F_1,F_2\}=} {}+\langle \nabla_P F_1,\phi(\nabla_M F_2)\cdot P\,\rangle-
\langle \nabla_P F_2,\phi(\nabla_M F_1)\cdot P\,\rangle\label{PB left red}
\end{gather}
for two arbitrary functions
$F_{1,2}(M,P):{\mathfrak g}^*\times O_p\mapsto{\mathbb R}$.
\end{theorem}

\begin{remark}
The formula (\ref{PB left red}) defines a Poisson bracket
not only on ${\mathfrak g}^*\times O_p$, but on all of ${\mathfrak g}^*\times V$. Rewriting
this formula as
\begin{gather*}
\{F_1,F_2\}=\langle M,[\nabla_M F_1,\nabla_M F_2]\,\rangle\nonumber\\
\phantom{\{F_1,F_2\}=} {}+\langle  P, \phi^*(\nabla_M F_2)\cdot \nabla_P F_1-
\phi^*(\nabla_M F_1)\cdot \nabla_P F_2\,\rangle%\label{PB left red once more}
\end{gather*}
one immediately identifies this bracket with the Lie--Poisson bracket of the
semiproduct Lie algebra ${\mathfrak g}\ltimes V^*$ corresponding to the representation
$-\phi^*$ of ${\mathfrak g}$ in $V^*$.
\end{remark}

We shall also need a version of the above theorem for the case when the
Lagrange function is invariant with respect to the {\it right action} of
some isotropy subgroup $G^{[A]}$ ($A\in V$), i.e.
\begin{equation*}%\label{right action}
{\mathbf L}(gh,R_{h*}\dot{g})={\mathbf L}(g,\dot{g}), \qquad h\in G^{[A]}.
\end{equation*}
In this case it is convenient to consider the right
trivialization of the tangent bundle:
\begin{equation*}%\label{right triv}
(g,\dot{g})\in TG\;\mapsto\;(g,\omega)\in G\times{\mathfrak g},
\end{equation*}
where
\begin{equation*}%\label{omega}
\omega=R_{g^{-1}*}\dot{g} \quad\Leftrightarrow\quad \dot{g}=R_{g*}\omega.
\end{equation*}
Denote the Lagrange function pushed through the trivialization map by
${\mathbf L}^{(r)}(g,\omega): G\times{\mathfrak g}\mapsto{\mathbb R}$, so that
\begin{equation*}%\label{Lagr right}
{\mathbf L}^{(r)}(g,\omega)={\mathbf L}(g,\dot{g}).
\end{equation*}
The corresponding invariance property of the Lagrange function is expressed as
\begin{equation*}%\label{right action for red}
{\mathbf L}^{(r)}(gh,\omega)={\mathbf L}^{(r)}(g,\omega), \qquad h\in G^{[A]}.
\end{equation*}
The formulas for the reduction of the Euler--Lagrange equations are also
slightly modified. As a section $(G\times{\mathfrak g})/G^{[A]}$ we choose the set
${\mathfrak g}\times O_{A}$, the reduction map being
\begin{equation*}
(g,\omega)\in G\times{\mathfrak g}\;\mapsto\; (\omega,a)\in {\mathfrak g}\times O_A, \qquad
{\rm where}  \quad a=\Phi(g)\cdot A\;,
\end{equation*}
so that the reduced Lagrange function ${\mathcal L}^{(r)}:{\mathfrak g}\times O_A
\mapsto{\mathbb R}$ is defined as
\begin{equation*}%\label{right red Lagr}
{\mathcal L}^{(r)}(\omega,a)={\mathbf L}^{(r)}(g,\omega),\qquad{\rm where}\quad
 a=\Phi(g)\cdot A.
\end{equation*}
There holds an analog of Theorem~\ref{continuous reduction left},
stating, in particular, that under the
right trivialization $(g,\dot{g})\mapsto(g,\omega)$ and the
subsequent reduction $(g,\omega)\mapsto(\omega,a)$, the Euler--Lagrange
equations {\rm (\ref{EL gen})} become the following {\bf right
Euler--Poincar\'e equations}:
\begin{equation}\label{EL right Ham red}
\left\{\begin{array}{l}
\dot{m}=-{\rm ad}^*\,\omega\cdot m-\nabla_a{\mathcal L}^{(r)}\diamond a,\vspace{1mm}\\
\dot{a}=\phi(\omega)\cdot a,\end{array}\right.
\end{equation}
where
\begin{equation*}%\label{m thru L red}
m=R_g^*\Pi=\nabla_{\omega}{\mathcal L}^{(r)}.
\end{equation*}

\renewcommand{\footnoterule}{\vspace*{3pt}%
\noindent
\rule{.4\columnwidth}{0.4pt}\vspace*{6pt}}

\section{Lagrangian mechanics and Lagrangian
reduction on \mbox{\mathversion{bold}$G\!\times \!G$}}

We now turn to the discrete time analog of these constructions.
Our presentation of the general discrete time Lagrangian mechanics
is an adaptation of the Moser--Veselov construction \cite{V,MV} for the case
when the basic manifold is a Lie group. The presentation of the discrete time
Lagrangian reduction follows \cite{BS1,BS2}. Almost all constructions and results
of the continuous time Lagrangian mechanics have their discrete time analogs.
The only exception is the existence of the ``energy'' integral (\ref{Ham gen}).

Let ${\mathbb L}(g,\hat{g}):G\times G\mapsto{\mathbb R}$ be a smooth function,
called the (discrete time) {\it Lagrange function}. For an arbitrary sequence
$\{g_k\in G,\;k=k_0,k_0+1,\ldots,k_1\}$ one can consider the {\it action
functional}
\begin{equation}\label{dS}
{\mathbb S}=\sum_{k=k_0}^{k_1-1}{\mathbb L}(g_k,g_{k+1}).
\end{equation}
Obviously, the sequences $\{g_k\}$ delivering extrema of this functional
(in the class of variations preserving $g_{k_0}$ and $g_{k_1}$), satisfy
with necessity the {\it discrete Euler--Lagrange equations}\footnote{For the
notations from the Lie groups theory used in this and
subsequent sections see, e.g., \cite{BS1}. In particular, for an arbitrary
smooth function $f:G\mapsto{\mathbb R}$, its right Lie derivative $d\,'f$ and
left Lie derivative $df$ are functions from $G$ into ${\mathfrak g}^*$ defined via the
formulas
\[
\langle df(g),\eta\rangle=\left.\frac{d}{d\epsilon}\,f\left(e^{\epsilon\eta}g\right)
\right|_{\epsilon=0},\qquad
\langle d\,'f(g),\eta\rangle=\left.\frac{d}{d\epsilon}\,f\left(ge^{\epsilon\eta}\right)
\right|_{\epsilon=0},\qquad \forall\; \eta\in{\mathfrak g},
\]
and the gradient $\nabla f(g)\in T^*_gG$ is defined as
\[
\nabla f(g)=R^*_{g^{-1}}\,df(g)=L^*_{g^{-1}}\,d\,'f(g).
\]}:
\begin{equation}\label{dEL}
\nabla_1{\mathbb L}(g_k,g_{k+1})+\nabla_2{\mathbb L}(g_{k-1},g_k)=0.
\end{equation}
Here $\nabla_1{\mathbb L}(g,\hat{g})$ ($\nabla_2{\mathbb L}(g,\hat{g})$) denotes the
gradient of ${\mathbb L}(g,\hat{g})$ with respect to the first argument $g$
(resp. the second argument $\hat{g}$). So, in our case, when $G$ is a Lie
group and not just a general smooth manifold, the equation (\ref{dEL}) is
written in a coordinate free form, using the intrinsic notions of the Lie
theory. As pointed out above, an invariant formulation of the Euler--Lagrange
equations in the continuous time case is more sophisticated. This seems
to underline the fundamental character of discrete Euler--Lagrange equations.

The equation (\ref{dEL}) is an implicit equation for $g_{k+1}$. In
general, it has more than one solution, and therefore defines a correspondence
(multi-valued map)  $(g_{k-1},g_k)\mapsto(g_k,g_{k+1})$. To discuss symplectic
properties of this correspondence, one defines:
\begin{equation*}%\label{dPi}
\Pi_k=\nabla_2{\mathbb L}(g_{k-1},g_k)\in T^*_{g_k}G.
\end{equation*}
Then (\ref{dEL}) may be rewritten as the following system:
\begin{equation}\label{dEL syst}
\left\{\begin{array}{l}
\Pi_k=-\nabla_1{\mathbb L}(g_k,g_{k+1}), \vspace{1mm}\\
\Pi_{k+1}=\nabla_2{\mathbb L}(g_k,g_{k+1}).
\end{array}\right.
\end{equation}
This system defines a (multivalued) map $(g_k,\Pi_k)\mapsto(g_{k+1},\Pi_{k+1})$
of $T^*G$ into itself. More precisely, the first equation in (\ref{dEL syst})
is an implicit equation for $g_{k+1}$, while the second one allows for the
explicit and unique calculation of $\Pi_{k+1}$, knowing $g_k$ and $g_{k+1}$. As
demonstrated in \cite{V,MV}, this map $T^*G\mapsto T^*G$ is symplectic
with respect to the standard symplectic structure on $T^*G$.

The tangent bundle $TG$ does not appear in the discrete time context at all.
On the contrary, the cotangent bundle $T^*G$ still plays an important role in
the discrete time theory, as the phase space with the canonical invariant
symplectic structure. The left trivialization of $T^*G$ is same as in the
continuous time case:
\begin{equation*}%\label{d left triv *}
(g_k,\Pi_k)\in T^*G \;\mapsto\; (g_k,M_k)\in G\times{\mathfrak g}^*,
\end{equation*}
where
\begin{equation*}%\label{d M}
M_k=L_{g_k}^*\Pi_k \quad\Leftrightarrow\quad \Pi_k=L_{g_k^{-1}}^* M_k.
\end{equation*}
Consider also the map
\begin{equation}\label{d left triv}
(g_k,g_{k+1})\in G\times G \;\mapsto\; (g_k,W_k)\in G\times G,
\end{equation}
where
\begin{equation*}%\label{W}
W_k=g_k^{-1}g_{k+1} \quad\Leftrightarrow\quad g_{k+1}=g_kW_k.
\end{equation*}
Denote the Lagrange function pushed through (\ref{d left triv}) by
\begin{equation*}%\label{dLagr left}
{\mathbb L}^{(l)}(g_k,W_k)={\mathbb L}(g_k,g_{k+1}).
\end{equation*}

Suppose that the Lagrange function ${\mathbb L}(g,\hat{g})$ is invariant under
the action of $G^{[p]}$ on $G\times G$ induced by left translations on $G$:
\begin{equation*}%\label{d left action}
{\mathbb L}(hg,h\hat{g})={\mathbb L}(g,\hat{g}), \qquad h\in G^{[p]}.
\end{equation*}
The corresponding invariance property of ${\mathbb L}^{(l)}(g,W)$ is expressed as:
\begin{equation*}%\label{d left action for red}
{\mathbb L}^{(l)}(hg,W)={\mathbb L}^{(l)}(g,W), \qquad h\in G^{[p]}.
\end{equation*}
We want to reduce the Euler--Lagrange equations with respect to this left
action. As a section $(G\times G)/G^{[p]}$ we choose the set $G\times O_{p}$.
The reduction map is
\begin{equation*}
(g,W)\in G\times G\;\mapsto\; (W,P)\in G\times O_p, \qquad {\rm where}
\quad P=\Phi(g^{-1})\cdot p,
\end{equation*}
so that the reduced Lagrange function $\Lambda^{(l)}: G\times O_p
\mapsto{\mathbb R}$ is defined as
\begin{equation*}%\label{left red dLagr}
\Lambda^{(l)}(W,P)={\mathbb L}^{(l)}(g,W),\qquad{\rm where}\quad
 P=\Phi(g^{-1})\cdot p.
\end{equation*}

\begin{theorem}\label{discrete semidirect reduction left} {\rm \cite{BS1,BS2}}

{\rm a)} Under the left trivialization $(g,\hat{g})\mapsto(g,W)$ and the
subsequent reduction $(g,W)\mapsto(W,P)$, the Euler--Lagrange
equations (\ref{dEL}) become the following {\bf left discrete time
Euler--Poincar\'e equations}:
\begin{equation}\label{dEL left Ham red}
\left\{\begin{array}{l}
{\rm Ad}^*\,W_k^{-1}\cdot M_{k+1}=M_k+\nabla_P
\Lambda^{(l)}(W_k,P_k)\diamond P_k,\vspace{1mm}\\
P_{k+1}=\Phi(W_k^{-1})\cdot P_k,\end{array}\right.
\end{equation}
where
\begin{equation*}%\label{d M thru L red}
M_k=d\,'_{\!W}\Lambda^{(l)}(W_{k-1},P_{k-1})\in{\mathfrak g}^*.
\end{equation*}

{\rm b)} If the ``Legendre transformation''
\begin{equation*}%\label{dLegendre left red}
(W_{k-1},P_{k-1})\in G\times O_p\mapsto
(M_k,P_k)\in {\mathfrak g}^*\times O_p,
\end{equation*}
where $P_k=\Phi(W_{k-1}^{-1})\cdot P_{k-1}$, is invertible, then
{\rm(\ref{dEL left Ham red})} define a map $(M_k,P_k)\mapsto(M_{k+1},P_{k+1})$
of  ${\mathfrak g}^*\times O_p$ which is Poisson with respect to the Poisson
bracket (\ref{PB left red}).
\end{theorem}

We shall also need a version of the above theorem for the case when the
Lagrange function is invariant with respect to the {\it right action} of
some isotropy subgroup $G^{[A]}$ ($A\in V$), i.e.
\begin{equation*}%\label{discr right action}
{\mathbb L}(gh,\hat{g}h)={\mathbb L}(g,\hat{g}), \qquad h\in G^{[A]}.
\end{equation*}
In this case it is convenient to consider the following map analogous to the
right trivialization of the tangent bundle:
\begin{equation*}%\label{discr right triv}
(g_k,g_{k+1})\in G\times G\;\mapsto\;(g_k,w_k)\in G\times G,
\end{equation*}
where
\begin{equation*}%\label{w}
w_k=g_{k+1}g_k^{-1} \quad\Leftrightarrow\quad g_{k+1}=w_kg_k.
\end{equation*}
Denote the Lagrange function pushed through this map by
${\mathbb L}^{(r)}(g_k,w_k): G\times G\mapsto{\mathbb R}$, so that
\begin{equation*}%\label{discr Lagr right}
{\mathbb L}^{(r)}(g_k,w_k)={\mathbb L}(g_k,g_{k+1}).
\end{equation*}
The corresponding invariance property of the Lagrange function is expressed as
\begin{equation*}
{\mathbb L}^{(r)}(gh,w)={\mathbb L}^{(r)}(g,w), \qquad h\in G^{[A]}.
\end{equation*}
The formulas for the reduction of the Euler--Lagrange equations are also
slightly modified. As a section $(G\times{\mathfrak g})/G^{[A]}$ we choose the set
${\mathfrak g}\times O_{A}$, the reduction map being
\begin{equation*}
(g,w)\in G\times G\;\mapsto\; (w,a)\in G\times O_A, \qquad
{\rm where}  \quad a=\Phi(g)\cdot A,
\end{equation*}
so that the reduced Lagrange function $\Lambda^{(r)}: G\times O_A
\mapsto{\mathbb R}$ is defined as
\begin{equation*}%\label{right red discr Lagr}
\Lambda^{(r)}(w,a)={\mathbb L}^{(r)}(g,w),\qquad{\rm where}\quad
 a=\Phi(g)\cdot A.
\end{equation*}
There holds an analog of Theorem~\ref{discrete semidirect reduction left},
stating, in particular, that under the
right trivialization $(g_k,g_{k+1})\mapsto(g_k,w_k)$ and the
subsequent reduction $(g_k,w_k)\mapsto(w_k,a_k)$, the Euler--Lagrange
equations {\rm (\ref{dEL syst})} become the following {\bf right discrete time
Euler--Poincar\'e equations}:
\begin{equation}\label{dEL right Ham red}
\left\{\begin{array}{l}
{\rm Ad}^*\,w_k\cdot m_{k+1}=m_k-\nabla_a
\Lambda^{(r)}(w_k,a_k)\diamond a_k,\vspace{1mm}\\
a_{k+1}=\Phi(w_k)\cdot a_k,\end{array}\right.
\end{equation}
where
\begin{equation*}%\label{d m thru L red}
m_k=d_w\Lambda^{(r)}(w_{k-1},a_{k-1})\in{\mathfrak g}^*.
\end{equation*}

The relation between the continuous time and the discrete time equations
is established, if we set
\begin{gather*}
g_k=g,\qquad g_{k+1}=g+\epsilon\dot{g}+O\left(\epsilon^2\right),\qquad
{\mathbb L}(g_k,g_{k+1})=\epsilon{\mathbf L}(g,\dot{g})+O\left(\epsilon^2\right);\\
P_k=P,\qquad W_k=I+\epsilon\Omega+O\left(\epsilon^2\right),\qquad
\Lambda^{(l)}(W_k,P_k)=\epsilon{\mathcal L}^{(l)}(\Omega,P)+O\left(\epsilon^2\right);\\
a_k=a,\qquad w_k=I+\epsilon\omega+O\left(\epsilon^2\right),\qquad
\Lambda^{(r)}(w_k,a_k)=\epsilon{\mathcal L}^{(r)}(\omega,a)+O\left(\epsilon^2\right).
\end{gather*}

\section{A multi-dimensional heavy top}
The basic Lie group relevant for our main examples is
\[
G={\rm SO}(n),\qquad {\rm so\ that} \quad {\mathfrak g}={\rm so}(n)
\]
(the ``physical'' rigid body corresponds to $n=3$). The scalar product
on ${\mathfrak g}$ is defined as
\[
\langle \xi,\eta\rangle=-\frac{1}{2}\,{\rm tr}\,(\xi\eta), \qquad
\xi,\eta\in{\mathfrak g}.
\]
This scalar product is used also to identify ${\mathfrak g}^*$ with ${\mathfrak g}$, so that
the previous formula can be considered also as a pairing  between the elements
$\xi\in{\mathfrak g}$ and $\eta\in{\mathfrak g}^*$.

The group $G$ is a natural configuration space for problems related to
the rotation of a rigid body. Indeed, if $E(t)=\left(e_1(t),\ldots,e_n(t)\right)$
stands for the time evolution of a certain orthonormal frame firmly
attached to the rigid body (all $e_k\in{\mathbb R}^n$, $\langle
e_k,e_j\rangle=\delta_{kj}$), then
\[
E(t)=g^{-1}(t)E(0)\quad \Leftrightarrow\quad e_k(t)=g^{-1}(t)e_k(0)\qquad
(1\le k\le n),
\]
with some $g(t)\in G$. The Lagrange function of an arbitrary rigid body
rotating about a fixed point in a homogeneous field with a linear potential
\begin{equation*}%\label{mLT pot}
\varphi(x)=\langle p,x\rangle,
\end{equation*}
is equal to
\begin{equation}\label{mLT Lagr 1}
{\mathbf L}(g,\dot{g})=-\frac{1}{2}\,{\rm tr}\,(\Omega J\Omega)-\langle p,a\rangle,
\end{equation}
where
\begin{itemize}
\itemsep0mm
\item $\Omega=g^{-1}\dot{g}=g^{\rm T}\dot{g}$ is the angular velocity of the
rigid body in the body frame $E(t)$;
\item $J$ is a symmetric matrix (tensor of inertia of the rigid body); choosing
the frame $E(0)$ properly, we can assure this matrix to be {\it diagonal},
$J={\rm diag}\,(J_1,\ldots,J_n)$, which will be supposed from now on;
\item $p\in\mathbb R^n$ is the constant gravity vector, calculated
in the rest frame;
\item $a=a(t)=g(t)A\in\mathbb R^n$ is the vector pointing from the fixed point
to the center of mass of the rigid body, calculated in the rest frame;
$A$ is the same vector calculated in the moving frame (where it is constant).
\end{itemize}
Obviously, the function (\ref{mLT Lagr 1}) may be rewritten also as
\begin{equation}\label{mLT Lagr 2}
{\mathbf L}(g,\dot{g})=\frac{1}{2}\langle\Omega,{\mathcal J}(\Omega)\rangle-\langle P,A\rangle,
\end{equation}
where ${\mathcal J}:{\rm so}(n)\mapsto{\rm so}(n)$ is the symmetric operator acting
as
\begin{equation*}
{\mathcal J}(\Omega)=J\Omega+\Omega J,
\end{equation*}
and $P(t)=g^{-1}(t)p$ is the gravity vector $p$ calculated in the moving frame.
The Lagrange function (\ref{mLT Lagr 1}) is in the framework of Section~2, if
the following identifications are made:
\begin{itemize}
\itemsep0mm
\item $V=V^*=\mathbb R^n$ with the standard Euclidean scalar product;
\item The representation $\Phi$ of $G$ in $V$ is defined as
\[
\Phi(g)\cdot v=gv\qquad {\rm for}\quad g\in{\rm SO}(n), \ v\in \mathbb R^n.
\]
\item Therefore the representation $\phi$ of ${\mathfrak g}$ in $V$ is given by
\[
\phi(\xi)\cdot v=\xi v\qquad {\rm for}\quad \xi\in{\rm so}(n),\ v\in\mathbb R^n,
\]
while the anti-representation $\phi^*$ of ${\mathfrak g}$ in $V^*$ is given by
\[
\phi^*(\xi)\cdot y=-\xi y\qquad {\rm for}\quad \xi\in{\rm so}(n),\ y\in\mathbb R^n,
\]
\item Finally, the bilinear operation $\diamond:V^*\times V\mapsto {\mathfrak g}^*$
is given by
\[
y\diamond v=vy^{\rm T}-yv^{\rm T}=v\wedge y\qquad {\rm for}\quad  y,v\in\mathbb R^n.
\]
\end{itemize}
The Lagrange function (\ref{mLT Lagr 2}) is manifestly invariant under the
left action of the isotropy subgroup $G^{[p]}$. Now Theorem~\ref{continuous reduction left}
is applicable,
which delivers the following {\it equations of motion of a heavy top in the
moving frame}:
\begin{equation}\label{mLT left}
\left\{\begin{array}{l} \dot{M}=[M,\Omega]+A\wedge P,\vspace{1mm}\\
\dot{P}=-\Omega P,\end{array}\right.
\end{equation}
where
\begin{equation*}%\label{mLT M}
M=\nabla_{\Omega}{\mathcal L}^{(l)}={\mathcal J}(\Omega)\quad \Leftrightarrow\quad
M_{jk}=(J_j+J_k)\Omega_{jk}.
\end{equation*}
According to the general theory, the system (\ref{mLT left}) is Hamiltonian
on the dual of the semidirect product Lie algebra
${\rm e}(n)={\rm so}(n)\ltimes\mathbb R^n$, with the Hamilton function
\begin{equation*}%\label{mLT Ham}
H(M,P)=\frac{1}{2}\langle M,\Omega\rangle+\langle P,A\rangle=
\frac{1}{2}\sum_{j<k} \frac{M_{jk}^2}{J_j+J_k}+\sum_{k=1}^nP_kA_k.
\end{equation*}
The correspondent invariant Poisson bracket reads:
\begin{gather}
\{M_{ij},M_{k\ell}\}  =  M_{i\ell}\delta_{jk}-M_{kj}\delta_{\ell i}-
M_{ik}\delta_{j\ell}+M_{\ell j}\delta_{ki},  \label{en LP br 1}\\
\{M_{ij},P_{k}\}  =  P_{i}\delta_{jk}-P_{j}\delta_{ik}.\label{en LP br 2}
\end{gather}

\section{The multi-dimensional Lagrange top}

The {\it multi-dimensional Lagrange top} is characterized by the following data:
$J_1=J_2=\cdots=J_{n-1}$, which means that the body is rotationally symmetric
with respect to the $n$th coordinate axis, and $A_1=A_2=\cdots=A_{n-1}=0$, which
means that the fixed point lies on the symmetry axis. Choosing units properly,
we may assume that
\begin{equation*}%\label{mLT data}
J_1=J_2=\cdots=J_{n-1}=\frac{\alpha}{2},\qquad J_n=1-\frac{\alpha}{2},\qquad
A=(0,0,\ldots,0,1)^{\rm T}.
\end{equation*}
The action of the operator ${\mathcal J}$ is given by
\begin{equation*}%\label{mLT operator}
M_{ij}={\mathcal J}(\Omega)_{ij}=\left\{\begin{array}{ll}
\alpha\Omega_{ij}, & 1\le i,j\le n-1,  \vspace{1mm}\\
\Omega_{ij}, & i=n\ {\rm or}\ j=n,\end{array}\right.
\end{equation*}
or in a more invariant fashion:
\begin{gather}
M={\mathcal J}(\Omega)  =  \alpha\Omega+(1-\alpha)\left(\Omega AA^{\rm T}+
AA^{\rm T}\Omega\right)   \label{mLT operator inv1}\\
 \phantom{M={\mathcal J}(\Omega)}{}=  \alpha\Omega-(1-\alpha)A\wedge(\Omega A).
                    \label{mLT operator inv2}
\end{gather}
Therefore, the formula (\ref{mLT Lagr 2}) may be represented as
\begin{equation}\label{mLT Lagr 3}
{\mathbf L}(g,\dot{g})=\frac{\alpha}{2}\langle\Omega,\Omega\rangle+
\frac{1-\alpha}{2}\langle \Omega A,\Omega A\rangle-\langle P,A\rangle.
\end{equation}

The ``Legendre transformation'' (\ref{mLT operator inv1}),
(\ref{mLT operator inv2}) is easily invertible. Notice that from
(\ref{mLT operator inv1}) there follows immediately
\begin{equation}\label{mTL Legendre inv interm}
MA=\Omega A,
\end{equation}
and plugging this into the second term of the right-hand side
of (\ref{mLT operator inv2}), we find:
\begin{equation*}%\label{mTL Legendre inv}
\Omega=\frac{1}{\alpha}\,M+\frac{1-\alpha}{\alpha}\,A\wedge(MA).
\end{equation*}

The integrability of the multi-dimensional Lagrange top was demonstrated by
Be\-lya\-ev~\cite{Be}, who constructed ``by hands'' the necessary number of
involutive integrals. A somewhat easier proof is delivered by the Lax
representation in the loop algebra ${\rm sl}(n+1)[\lambda,\lambda^{-1}]$
twisted by the Cartan automorphism \cite{RS1,RS2}.
\begin{proposition}
For the Lagrange top, the moving frame equations (\ref{mLT left}) are
equivalent to the matrix equation
\begin{equation*}
\dot{L}(\lambda)=[L(\lambda),U(\lambda)],
\end{equation*}
where
\begin{equation*}
L(\lambda)=\left(\begin{array}{cc}
M & \lambda A-\lambda^{-1} P \vspace{1mm}\\ \lambda A^{\rm T}-\lambda^{-1}P^{\rm T} & 0
\end{array}\right),\qquad
U(\lambda)=\left(\begin{array}{cc}
\Omega & \lambda A \vspace{1mm}\\ \lambda A^{\rm T} & 0
\end{array}\right).
\end{equation*}
\end{proposition}

\begin{proof} A direct calculation based on the formula
(\ref{mTL Legendre inv interm}).
\end{proof}

Integrability is not the only distinctive feature of the Lagrange top.
Another one is the existence of a nice Euler--Poincar\'e description not
only in the moving frame, but also in the rest one.
Rewriting (\ref{mLT Lagr 3}) as
\begin{equation}\label{mLT Lagr 4}
{\mathbf L}(g,\dot{g})=\frac{\alpha}{2}\langle\omega,\omega\rangle+
\frac{1-\alpha}{2}\langle \omega a,\omega a\rangle-\langle p,a\rangle,
\end{equation}
where $\omega=\dot{g}g^{-1}$, we observe that the Lagrange function
of the Lagrange top is not only left-invariant (with respect to $G^{[p]}$),
but also right-invariant (with respect to $G^{[A]}$). Therefore, we may
apply the formula (\ref{EL right Ham red}), which in the present setup reads:
\begin{equation}\label{mLT right Ham red}
\left\{\begin{array}{l}
\dot{m}=[\omega,m]+\nabla_a{\mathcal L}^{(r)}\wedge a,\vspace{1mm}\\
\dot{a}=\omega a,\end{array}\right.
\end{equation}
Straightforward calculations based on (\ref{mLT Lagr 4}) give:
\begin{gather*}
\nabla_a{\mathcal L}^{(r)}=-(1-\alpha)\omega^2a-p,\\
m=\nabla_{\omega}{\mathcal L}^{(r)}=\alpha\omega+(1-\alpha)\left(\omega aa^{\rm T}+
aa^{\rm T}\omega\right).
\end{gather*}
The last formula implies, first,
\[
ma=\alpha\omega a+(1-\alpha)\omega a\langle a,a\rangle=\omega a,
\]
and, second,
\[
[\omega,m]-(1-\alpha)\left(\omega^2a\right)\wedge a=\left[\omega,m-(1-\alpha)
\left(\omega aa^{\rm T}+aa^{\rm T}\omega\right)\right]=[\omega,\alpha\omega]=0.
\]
Plugging these results into (\ref{mLT right Ham red}), we finally arrive at
the following nice system:
\begin{equation}\label{mLT right}
\left\{\begin{array}{l}
\dot{m}=a\wedge p,\vspace{1mm}\\
\dot{a}=m a,\end{array}\right.
\end{equation}
where $m=gMg^{-1}$ is the kinetic moment in the rest frame. This is a
Hamiltonian system with respect to the (minus) Lie--Poisson bracket of
${\rm e}(n)$:
\begin{gather}
\{m_{ij},m_{k\ell}\}  =  -m_{i\ell}\delta_{jk}+m_{kj}\delta_{\ell i}+
m_{ik}\delta_{j\ell}-m_{\ell j}\delta_{ki},  \label{en LP r br 1}\\
\{m_{ij},a_{k}\}  =  -a_{i}\delta_{jk}+a_{j}\delta_{ik},\label{en LP r br 2}
\end{gather}
with the Hamilton function
\begin{equation*}%\label{mLT right H}
H(m,a)=\frac{1}{2}\langle m,m\rangle+\langle a,p\rangle.
\end{equation*}
A remarkable feature of the system (\ref{mLT right}) is its independence on
the anisotropy parameter~$\alpha$.
\begin{proposition}
The rest frame equations (\ref{mLT right}) of the Lagrange top are equivalent
to the matrix equation
\begin{equation*}
\dot{\ell}(\lambda)=[\ell(\lambda),u(\lambda)],
\end{equation*}
where
\begin{equation*}
\ell(\lambda)=\left(\begin{array}{cc}
m & \lambda a-\lambda^{-1} p \vspace{1mm}\\
\lambda a^{\rm T}-\lambda^{-1}p^{\rm T} & 0
\end{array}\right),\qquad
u(\lambda)=\left(\begin{array}{cc}
0 & \lambda a \vspace{1mm}\\ \lambda a^{\rm T} & 0  \end{array}\right).
\end{equation*}
\end{proposition}

\section{A discrete time analog of the Lagrange top}

An integrable discretization of the 3-dimensional Lagrange top was constructed
in~\cite{BS1}, where Lagrangian dynamics on $G={\rm SU}(2)$ are used, leading
to Euler--Poincar\'e equations on ${\rm su}(2)\ltimes {\rm su}(2)$.
The multi-dimensional generalization presented here deals
with Lagrangians on $G={\rm SO}(n)$, leading to Euler--Poincar\'e equations
on ${\rm so}(n)\ltimes\mathbb R^n$. This change of context (when compared with
\cite{BS1}) required a modification of the kinetic energy terms. Of course,
this is essential only in the discrete time setting, since in the continuous
time situation the kinetic energy is given purely in terms of the Lie algebra
${\mathfrak g}$ of the Lie group $G$, and the Lie algebras ${\rm su}(2)$ and
${\rm so}(3)$ are isomorphic.

\subsection{Rest frame formulation}
Consider the following discrete analog of the Lagrange function
(\ref{mLT Lagr 4}):
\begin{gather}
{\mathbb L}(g_k,g_{k+1})=
-\frac{\alpha}{\epsilon}\,{\rm tr}\log
\left(2I+w_k+w_k^{-1}\right)\nonumber\\
\phantom{{\mathbb L}(g_k,g_{k+1})=} {}-\frac{2(1-\alpha)}{\epsilon}\,\log\left(1+\langle
a_k,a_{k+1}\rangle\right)-\epsilon\langle p,a_k\rangle.\label{dmLT Lagr 1}
\end{gather}
where $w_k$, $a_k$ are defined as in Section~3:
\[
w_k=g_{k+1}g_k^{-1},\qquad a_k=g_kA,\qquad{\rm so\ that}\quad a_{k+1}=w_ka_k.
\]
The powers of $\epsilon$ are introduced in a way assuring the correct
asymptotics in the continuous limit $\epsilon\to 0$, namely
${\mathbb L}(g_k,g_{k+1})\approx \epsilon{\mathbf L}(g,\dot{g})$, as $g_k=g$ and
$g_{k+1}\approx g+\epsilon\dot{g}$ (see the end of Section~3). This is seen
with the help of the following simple lemma.
\begin{lemma}\label{discrete Lagrange lemma}
Let $w(\epsilon)=I+\epsilon\omega+O\left(\epsilon^2\right)\in {\rm SO}(n)$ be a smooth
curve, $\omega\in {\rm so}(n)$. Then
\begin{equation}\label{tr as}
{\rm tr}\log\left(2I+w(\epsilon)+w^{-1}(\epsilon)\right)=
{\rm const}-\frac{\epsilon^2}{2}\,\langle\omega,\omega\rangle+O\left(\epsilon^3\right).
\end{equation}
For an arbitary $a\in\mathbb R^n$:
\begin{equation}\label{scalpr as}
\langle a,w(\epsilon)a\rangle=\langle a,a\rangle-
\frac{\epsilon^2}{2}\,\langle \omega a,\omega a\rangle+O\left(\epsilon^3\right).
\end{equation}
\end{lemma}

\begin{proof} Let $w=I+\epsilon\omega+\epsilon^2 v+O\left(\epsilon^3\right)$.
Then from $ww^{\rm T}=I$ we get:
\begin{equation}\label{lemma aux}
v+v^{\rm T}+\omega\omega^{\rm T}=0\quad\Rightarrow\quad
v=\frac{1}{2}\,\omega^2+v_1,\quad v_1\in {\rm so}(n).
\end{equation}
Hence
\begin{gather*}
2I+w+w^{\rm T}=4I+\epsilon^2\omega^2+O\left(\epsilon^3\right)\\
\qquad \Rightarrow\quad
\log\left(2I+w+w^{\rm T}\right)={\rm const}\,I+
\frac{\epsilon^2}{4}\,\omega^2+O\left(\epsilon^3\right),
\end{gather*}
which yields (\ref{tr as}). Similarly, we derive from (\ref{lemma aux}):
\[
a^{\rm T}wa=a^{\rm T}a+\frac{\epsilon^2}{2}a^{\rm T}\omega^2a+O\left(\epsilon^3\right),
\]
which implies (\ref{scalpr as}).
\end{proof}

Writing in (\ref{dmLT Lagr 1}) $w_ka_k$ for $a_{k+1}$, we come to the Lagrange
function $\Lambda^{(r)}(a_k,w_k)={\mathbb L}(g_k,g_{k+1})$ depending only on $a_k$,
$w_k$, hence invariant with respect to the right action of $G^{[A]}$. It may
be reduced following the procedure of Section~3.

\begin{theorem}\label{discrete mLagrange top in rest frame}
The Euler--Lagrange equations of motion for the Lagrange function
(\ref{dmLT Lagr 1}) are equivalent to the following system:
\begin{equation}\label{dLT rest frame}
\left\{\begin{array}{l}
m_{k+1}=m_k+\epsilon a_k\wedge p,\vspace{2mm}\\
a_{k+1}=\displaystyle\frac{I+(\epsilon/2) m_{k+1}}{I-(\epsilon/2)m_{k+1}}\,a_k.
\end{array}\right.
\end{equation}
The map $(m_k,a_k)\mapsto(m_{k+1},a_{k+1})$ is Poisson with respect to the
bracket (\ref{en LP r br 1}), (\ref{en LP r br 2}), and the following function
is its integral of motion:
\begin{equation}\label{dLT rest H}
H_{\epsilon}(m,a)=\frac{1}{2}\langle m,m\rangle +\langle a,p\rangle
-\frac{\epsilon}{2}\langle m,p\wedge a\rangle.
\end{equation}
\end{theorem}

\begin{proof} The right reduced Euler--Lagrange equations
(\ref{dEL right Ham red}) have in the present setup the following form:
\begin{equation}\label{Right inv dLagr}
\left\{\begin{array}{l}
w_k^{-1}m_{k+1}w_k=m_k-a_k\wedge \nabla_a \Lambda^{(r)}(a_k,w_k),
\vspace{1mm}\\
a_{k+1}=w_ka_k,\end{array}\right.
\end{equation}
where
\begin{equation*}%\label{dLT m}
m_{k+1}=d_w\Lambda^{(r)}(a_k,w_k).
\end{equation*}
To calculate the derivatives of $\Lambda^{(r)}$, we use the following formulas:
\begin{gather}\label{deriv aux1}
d_w\,{\rm tr}\log\left(2I+w_k+w_k^{-1}\right)=-2\,\frac{w_k-I}{w_k+I},\\
\label{deriv aux2}
d_w\langle a_k,w_ka_k\rangle=a_k\wedge (w_ka_k),\qquad
\nabla_a\langle a_k,w_ka_k\rangle=w_ka_k+w_k^{\rm T}a_k.
\end{gather}
Indeed, the first one of these expressions follows from:
\begin{gather*}
\left\langle d_w\,{\rm tr}\log\left(2I+w_k+w_k^{-1}\right),\eta\right\rangle  =
\left.\frac{d}{d\varepsilon}\,{\rm tr}\log\left(2I+e^{\varepsilon\eta}w_k+
w_k^{-1}e^{-\varepsilon\eta}\right)\right|_{\varepsilon=0}\\
\qquad  =  {\rm tr}\left(\left(2I+w_k+w_k^{-1}\right)\left(w_k-w_k^{-1}\right)\eta\right)
 =  {\rm tr}\left((w_k+I)^{-1}(w_k-I)\eta\right)\\
\qquad =  -2\left\langle (w_k+I)^{-1}(w_k-I),\eta\right\rangle.
\end{gather*}
To prove the second one, proceed similarly:
\begin{gather*}
\left\langle d_w\langle a_k,w_ka_k\rangle,\eta\right\rangle  =
\left.\frac{d}{d\varepsilon}\langle a_k, e^{\varepsilon\eta}w_ka_k
\rangle\right|_{\varepsilon=0}=\langle a_k,\eta w_ka_k\rangle\\
 \qquad  =  {\rm tr}\left(w_ka_ka_k^{\rm T}\eta\right)=
 \frac{1}{2}\,{\rm tr}\left(\left(w_ka_ka_k^{\rm T}-a_ka_k^{\rm T}w_k^{\rm T}\right)\eta\right)
 =  \left\langle a_k\wedge(w_ka_k),\eta\right\rangle.
\end{gather*}
Finally, for the third expression we have:
\[
\left\langle \nabla_a\langle a_k,w_ka_k\rangle,\eta\right\rangle =
\left.\frac{d}{d\varepsilon}\langle a_k+\varepsilon\eta,
w_k(a_k+\varepsilon\eta)\rangle\right|_{\varepsilon=0}
 =\left\langle w_ka_k+w_k^{\rm T}a_k,\eta\right\rangle.
\]

With the help of (\ref{deriv aux1}), (\ref{deriv aux2}) we find:
\begin{equation}\label{dLT m expr}
m_{k+1}=\frac{2\alpha}{\epsilon}\,\frac{w_k-I}{w_k+I}-
\frac{2(1-\alpha)}{\epsilon}\,\frac{a_k\wedge a_{k+1}}
{1+\langle a_k,a_{k+1}\rangle},
\end{equation}
and
\begin{gather*}
w_k^{-1}m_{k+1}w_k+a_k\wedge\nabla_a\Lambda^{(r)}(a_k,w_k)
\nonumber\\
\qquad  =  \frac{2\alpha}{\epsilon}\,\frac{w_k-I}{w_k+I}-
\frac{2(1-\alpha)}{\epsilon}\,
\frac{a_k\wedge a_{k+1}}{1+\langle a_k,a_{k+1}\rangle}-\epsilon a_k\wedge p
 =  m_{k+1}-\epsilon a_k\wedge p.
\end{gather*}
Comparing the latter formula with the first equation of motion in
(\ref{Right inv dLagr}), we find that it can be rewritten as
\[
m_k=m_{k+1}-\epsilon a_k\wedge p,
\]
which is the first equation in (\ref{dLT rest frame}).

To obtain the second one, derive from (\ref{dLT m expr}):
\begin{gather*}
\frac{\epsilon}{2}m_{k+1}\,a_{k+1}  =
\alpha\,\frac{w_k-I}{w_k+I}\,a_{k+1}-(1-\alpha)
\frac{a_k-a_{k+1}\langle a_k,a_{k+1}\rangle}{1+\langle a_k,a_{k+1}\rangle},\\
\frac{\epsilon}{2}m_{k+1}\,a_k  =
\alpha\,\frac{w_k-I}{w_k+I}\,a_k-(1-\alpha)\frac{a_k\langle
a_k,a_{k+1}\rangle-a_{k+1}}{1+\langle a_k,a_{k+1}\rangle}.
\end{gather*}
Adding these two equations, we find:
\[
\frac{\epsilon}{2}m_{k+1}(a_{k+1}+a_k)=
\alpha\,\frac{w_k-I}{w_k+I}(a_{k+1}+a_k)+(1-\alpha)(a_{k+1}-a_k)=
a_{k+1}-a_k.
\]
(On the last step we took into account that $a_{k+1}=w_ka_k$.)
This is nothing but the second equation of motion in (\ref{dLT rest frame}).

The Poisson properties of the map (\ref{dLT rest frame}) are assured
by the version of Theorem~\ref{discrete semidirect reduction left} for
right-invariant Lagrangians.

It remains to demonstrate that the function (\ref{dLT rest H}) is indeed an
integral of motion. This is done by the following derivation:
\begin{gather*}
H_{\epsilon}(m_{k+1},a_{k+1})  =
\frac{1}{2}\langle m_{k+1},m_{k+1}\rangle +\langle a_{k+1}-\frac{\epsilon}{2}\,
m_{k+1}a_{k+1},p\,\rangle \\
 \phantom{H_{\epsilon}(m_{k+1},a_{k+1})} =
\frac{1}{2}\langle m_{k+1},m_{k+1}\rangle +\langle a_k+\frac{\epsilon}{2}\,
m_{k+1}a_k,p\,\rangle\\
 \phantom{H_{\epsilon}(m_{k+1},a_{k+1})}=  \frac{1}{2}\langle m_{k+1},m_{k+1}+\epsilon p\wedge a_k\,\rangle
+\langle a_k,p\,\rangle\\
 \phantom{H_{\epsilon}(m_{k+1},a_{k+1})}=  \frac{1}{2}\langle m_k-\epsilon p\wedge a_k, m_k\rangle
+\langle a_k,p\,\rangle = H_{\epsilon}(m_k,a_k).
\end{gather*}
(In this calculation we used the identity $\langle ma,p\rangle=
\langle m,p\wedge a\rangle$ for $m\in{\rm so}(n)$, $p,a\in\mathbb R^n$;
notice that the scalar products on the both sides of this identity are
defined on different spaces!) The theorem is proved.
\end{proof}

Actually, the map (\ref{dLT rest frame}) possesses not only the integral
(\ref{dLT rest H}) but a full set of involutive integrals necessary for
the complete integrability. The most direct way to the proof of this
statement is, as usual, through the Lax representation which lives, just as
in the continuous time situation, in the loop algebra
${\rm sl}(n+1)[\lambda,\lambda^{-1}]$ twisted by the Cartan automorphism.
\begin{theorem}
The map (\ref{dLT rest frame}) admits the following Lax representation:
\begin{equation}\label{dmLT rest Lax}
\ell_{k+1}(\lambda)=v_k^{-1}(\lambda)\ell_k(\lambda)v_k(\lambda),
\end{equation}
with the matrices
\begin{gather*}
\ell_k(\lambda)=\left(\begin{array}{cc}
m_k & \lambda b_k-\lambda^{-1} p \vspace{1mm}\\
\lambda b_k^{\rm T}-\lambda^{-1}p^{\rm T} & 0
\end{array}\right), \qquad
v_k(\lambda)=\frac{I+(\epsilon/2)u_k(\lambda)}{I-(\epsilon/2)u_k(\lambda)},\nonumber\\
 u_k(\lambda)=\left(\begin{array}{cc}
0 & \lambda a_k \vspace{1mm}\\ \lambda a_k^{\rm T} & 0  \end{array}\right),
\end{gather*}
where the following abbreviation is used:
\begin{equation}\label{bk}
b_k=\left(I-\frac{\epsilon}{2}\,m_k\right)a_k+\frac{\epsilon^2}{4}\,p.
\end{equation}
\end{theorem}

\begin{proof} Direct verification. Notice that it is most convenient to
check (\ref{dmLT rest Lax}) in the form
\[
\left(I+\frac{\epsilon}{2}u_k(\lambda)\right)\ell_{k+1}(\lambda)
\left(I-\frac{\epsilon}{2}u_k(\lambda)\right)=
\left(I-\frac{\epsilon}{2}u_k(\lambda)\right)\ell_k(\lambda)
\left(I+\frac{\epsilon}{2}u_k(\lambda)\right),
\]
when it becomes polynomial in $\lambda$.
\end{proof}

This theorem provides us with a complete set of integrals of motion of our
discrete ti\-me Lagrangian map: these are the coefficients of the characteristic
polyno\-mial $\det\!\left(\ell_k(\lambda)\!-\mu I\right)$. Notice that these integrals
of motion {\it do not} coincide with the integrals of motion of the
continuous-time problem. To be more concrete, the integrals of motion of
the map (\ref{dLT rest frame}) are obtained from the integrals of the
continuous time problem by replacing $a$ by $b=a+O(\epsilon)$ given in
(\ref{bk}). As for the actual integration of our map in terms of
theta-functions, we leave it as an open problem for the
interested reader (cf., e.g., \cite{RvM,GZ}).

\subsection{Moving frame formulation}

It turns out that the equations of the discrete time Lagrange top in the
moving frame formulation become a bit nicer under a little
inessential change of the Lagrange function~(\ref{dmLT Lagr 1}), namely under
replacing $\langle p,a_k\rangle$ in the last term on the right-hand side by
$\langle p,a_{k+1}\rangle$. This modification does not influence the discrete
action functional (\ref{dS}), apart from the boundary terms. It is not
difficult to see that this modification is equivalent to exchanging
$w_k\leftrightarrow w_k^{-1}$, $a_k\leftrightarrow a_{k+1}$, which in turn is
equivalent to considering the evolution backwards in time with the
simultaneous change $\epsilon\mapsto -\epsilon$. Now express the discrete
Lagrange function (\ref{dmLT Lagr 1}), with the above modification, in terms of
$P_k=g_k^{-1}p$ and $W_k=g_k^{-1}g_{k+1}$:
\begin{gather}
{\mathbb L}(g_k,g_{k+1})=\Lambda^{(l)}(P_k,W_k)
  =  -\frac{\alpha}{\epsilon}\,{\rm tr}\log
\left(2I+W_k+W_k^{-1}\right)\nonumber\\
\qquad {}-\frac{2(1-\alpha)}{\epsilon}\,\log\left(1+\langle
A,W_kA\rangle\right)-\epsilon\langle P_k,W_kA\rangle.\label{dmLT Lagr body}
\end{gather}
Since $W_k=I+\epsilon\Omega+O\left(\epsilon^2\right)$, we can apply
Lemma~\ref{discrete Lagrange lemma} to see that
\[
\Lambda^{(l)}(P_k,W_k)=\epsilon{\mathcal L}^{(l)}(P,\Omega)+O\left(\epsilon^2\right),
\]
where ${\mathcal L}^{(l)}(P,\Omega)$ is the Lagrange function (\ref{mLT Lagr 2})
of the continuous time Lagrange top. Now, one can derive all results concerning
the discrete time Lagrange top in the body frame from the ones in the rest
frame by performing the change of frames so that
\[
M_k=g_k^{-1}m_kg_k,\qquad P_k=g_k^{-1}p,\qquad A=g_k^{-1}a_k,
\]
and taking into account the modification mentioned above. Alternatively, one
can derive them independently from and similarly to the rest frame results,
applying Theorem~\ref{discrete semidirect reduction left}, the main result
of which, the system (\ref{dEL left Ham red}), takes in the present setup the
form
\begin{equation}\label{dEL top left red}
\left\{\begin{array}{l}
W_kM_{k+1}W_k^{-1}=M_k+P_k\wedge\nabla_P\Lambda^{(l)}(W_k,P_k),\vspace{1mm}\\
P_{k+1}=W_k^{-1}P_k,\end{array}\right.
\end{equation}
where
\begin{equation}\label{dEL top left M}
M_{k+1}=d\,'_{\!W}\Lambda^{(l)}(W_k,P_k).
\end{equation}
Anyway, the corresponding results read:
\begin{theorem}\label{discrete mLagrange top in moving frame}
The Euler--Lagrange equations for the Lagrange function
(\ref{dmLT Lagr body}) are equivalent to the following system:
\begin{equation}\label{dmLT body}
\left\{\begin{array}{l}
M_{k+1}=W_k^{-1}M_kW_k+\epsilon A\wedge P_{k+1},\vspace{1mm}\\
P_{k+1}=W_k^{-1}P_k,
\end{array}\right.
\end{equation}
where the ``angular velocity'' $W_k\in{\rm SO}(n)$ is related to the
``angular momentum'' $M_k\in{\rm so}(n)$ via the ``Legendre transformation'':
\begin{equation}\label{dmLT W thru M}
M_k=\frac{2\alpha}{\epsilon}\cdot\frac{W_k-I}{W_k+I}-
\frac{2(1-\alpha)}{\epsilon}\cdot\frac{A\wedge(W_kA)}
{1+\langle A,W_kA\rangle}.
\end{equation}
The map (\ref{dmLT body}), (\ref{dmLT W thru M}) is Poisson with respect
to the Poisson bracket (\ref{en LP br 1}), (\ref{en LP br 2}), and has a
complete set of involutive integrals assuring its complete integrability.
One of these integrals is
\begin{equation*}%\label{dmLT body H}
\bar{H}_{\epsilon}(M,P)=\frac{1}{2}\langle M,M\rangle +\langle P,A\rangle
+\frac{\epsilon}{2}\langle M,P\wedge A\rangle.
\end{equation*}
\end{theorem}

\begin{proof} The derivation of (\ref{dmLT body}), (\ref{dmLT W thru M}) is
straightforward, like in Theorem~\ref{discrete mLagrange top in rest frame}.
Obviously, due to $W_k=I+\epsilon\Omega+O\left(\epsilon^2\right)$, the equations of motion
(\ref{dmLT body}) approximate the continuous time ones (\ref{mLT left}),
while the ``Legendre  transformation'' (\ref{dmLT W thru M}) approximates
(\ref{mLT operator inv2}). We discuss now the inversion of the ``Legendre
transformation'' (\ref{dmLT W thru M}). Obviously, it is trivially invertible
if $\alpha=1$: then
\[
M_k=\frac{2}{\epsilon}\cdot\frac{W_k-I}{W_k+I}\quad\Rightarrow\quad
W_k=\frac{I+(\epsilon/2)M_k}{I-(\epsilon/2)M_k}.
\]
For a general $\alpha$ we derive from (\ref{dmLT W thru M}):
\begin{equation}\label{dmTL Legendre inv interm}
W_kA=\frac{I+(\epsilon/2)M_k}{I-(\epsilon/2)M_k}\,A.
\end{equation}
Indeed, we have:
\begin{gather*}
\frac{\epsilon}{2}M_kA  =
\alpha\,\frac{W_k-I}{W_k+I}\,A-(1-\alpha)
\frac{A\langle A,W_kA\rangle-W_kA}{1+\langle A,W_kA\rangle},\\
\frac{\epsilon}{2}M_kW_kA  =
\alpha\,\frac{W_k-I}{W_k+I}\,W_kA-(1-\alpha)\frac{A-W_kA\langle
A,W_kA\rangle}{1+\langle A,W_kA\rangle}.
\end{gather*}
Adding these two equations, we find:
\[
\frac{\epsilon}{2}M_k(A+W_kA)=
\alpha\,\frac{W_k-I}{W_k+I}(A+W_kA)-(1-\alpha)(A-W_kA)=W_kA-A,
\]
which yields (\ref{dmTL Legendre inv interm}). Now plugging the expression
for $W_kA$ through $M_k$, $A$ into the second term on the right-hand side
of (\ref{dmLT W thru M}), we see that this equation is uniquely solvable for
$W_k$. Turning to the last statement of Theorem \ref{discrete mLagrange
top in moving frame}, we have:
\begin{gather*}
\bar{H}_{\epsilon}(M_{k+1},P_{k+1})  =
\frac{1}{2}\langle M_{k+1},M_{k+1}+\epsilon P_{k+1}\wedge A\rangle +
\langle P_{k+1},A\rangle \\
\phantom{\bar{H}_{\epsilon}(M_{k+1},P_{k+1})} =
\frac{1}{2}\langle W_k^{-1}M_kW_k-\epsilon P_{k+1}\wedge A,
W_k^{-1}M_kW_k\rangle +\langle P_{k+1},A\rangle\\
 \phantom{\bar{H}_{\epsilon}(M_{k+1},P_{k+1})}
=  \frac{1}{2}\langle M_k,M_k\rangle
+\langle P_{k+1},A-\frac{\epsilon}{2}\,W_k^{-1}M_kW_kA\rangle\\
\phantom{\bar{H}_{\epsilon}(M_{k+1},P_{k+1})} =  \frac{1}{2}\langle M_k,M_k\rangle
+\langle P_k,W_kA-\frac{\epsilon}{2}\,M_kW_kA\rangle\\
\phantom{\bar{H}_{\epsilon}(M_{k+1},P_{k+1})}=  \frac{1}{2}\langle M_k,M_k\rangle
+\langle P_k,A+\frac{\epsilon}{2}\,M_kA\rangle =
\bar{H}_{\epsilon}(M_k,P_k).
\end{gather*}
(On the last but one step we used the formula
(\ref{dmTL Legendre inv interm}).)
\end{proof}

We close this section with a Lax representation for the map (\ref{dmLT body}),
(\ref{dmLT W thru M}).
\begin{theorem}
The map (\ref{dmLT body}), (\ref{dmLT W thru M}) has the following Lax
representation:
\begin{equation*}%\label{dmLT body Lax}
L_{k+1}(\lambda)=V_k^{-1}(\lambda)L_k(\lambda)V_k(\lambda),
\end{equation*}
with the matrices
\begin{gather*}
L_k(\lambda)  =  \left(\begin{array}{cc}
M_k & \lambda B_k-\lambda^{-1} P_k \vspace{1mm}\\
\lambda B_k^{\rm T}-\lambda^{-1}P_k^{\rm T} & 0
\end{array}\right),\\%\label{dmLT body Lax L}\\
V_k(\lambda)  =  \left(\begin{array}{cc}
W_k & 0 \\ 0 & 1 \end{array}\right)\frac{I+(\epsilon/2)U(\lambda)}
{I-(\epsilon/2)U(\lambda)},\qquad U(\lambda)=
\left(\begin{array}{cc}
0 & \lambda A \\ -\lambda A^{\rm T} & 0
\end{array}\right),   %\label{dmLT body Lax U}
\end{gather*}
where the following abbreviation is used:
\begin{equation*}%\label{Bk}
B_k=\left(I+\frac{\epsilon}{2}\,M_k\right)A+\frac{\epsilon^2}{4}\,P_k.
\end{equation*}
\end{theorem}

\section{The Clebsch case of the rigid body motion\\
in an ideal fluid}

Clebsch~\cite{C} found an integrable case of the motion of a
3-dimensional rigid body in an ideal fluid, which was generalized
in~\cite{P} for the $n$-dimensional situation. This problem is traditionally
described by a Hamiltonian system on ${\rm e}^*(n)$ with the Hamilton function
\begin{equation*}%\label{Kirch H}
H(M,P)=\frac{1}{2}\sum_{j<k} c_{jk}M_{jk}^2-
\frac{1}{2}\sum_{k=1}^nb_kP_k^2.
\end{equation*}
Here $(M,P)\in{\rm e}^*(n)$, so that $M\in{\rm so}(n)$, $P\in\mathbb R^n$,
and $C=\{c_{jk}\}_{j,k=1}^n$, $B={\rm diag}\,(b_k)$ are some symmetric matrices.
The equations of motion read:
\begin{equation}\label{Kirch}
\left\{\begin{array}{l} \dot{M}=[M,\Omega]+P\wedge (BP),\vspace{1mm}\\
\dot{P}=-\Omega P,\end{array}\right.
\end{equation}
where the matrix $\Omega\in{\rm so}(n)$ is defined by the formula
\begin{equation*}%\label{Kirch Omega}
\Omega_{jk}=c_{jk}M_{jk}.
\end{equation*}
The ``physical'' Lagrangian formulation of this problem is dealing with
a Lagrangian on the group ${\rm E}(n)$, left-invariant under the action of
a whole group. However, it may be obtained also in the framework of
Section~2, from a Lagrangian on ${\rm SO}(n)$, left-invariant
under the action of an isotropy subgroup of some element $p\in\mathbb R^n$.
These two different settings lead to formally identical results.

So, one considers the Lagrange function
\begin{equation}\label{Kirch Lagr}
{\mathbf L}(g,\dot{g})={\mathcal L}^{(l)}(\Omega,P)=
\frac{1}{2}\,\langle\Omega,{\mathcal J}(\Omega)\rangle+
\frac{1}{2}\,\langle P,BP\rangle.
\end{equation}
Here $(g,\dot{g})\in T{\rm SO}(n)$, and $\Omega=g^{-1}\dot{g}\in{\rm so}(n)$,
$P=g^{-1}p\in\mathbb R^n$. It is supposed that ${\mathcal J}:{\rm so}(n)\mapsto
{\rm so}(n)$ is a linear operator acting as
\begin{equation*}%\label{Kirch cJ}
{\mathcal J}(\Omega)_{jk}=c_{jk}^{-1}\Omega_{jk}.
\end{equation*}
The reduced equations of motion for this Lagrange function delivered by
Theorem~\ref{continuous reduction left} coincide with (\ref{Kirch}).

The Clebsch case is characterized by the relations
\begin{equation*}%\label{Clebsch cond}
\frac{b_i-b_j}{c_{ij}}+\frac{b_j-b_k}{c_{jk}}+\frac{b_k-b_i}{c_{ki}}=0,
\end{equation*}
which implies that
\begin{equation*}%\label{Clebsch coefs}
c_{jk}=\frac{b_j-b_k}{a_j-a_k}
\end{equation*}
for some matrix $A={\rm diag}\,(a_k)$. The Lax representation found in~\cite{P}
reads:
\begin{equation*}%\label{Clebsch Lax}
\dot{L}(\lambda)=[L(\lambda),U(\lambda)],
\end{equation*}
where
\begin{equation*}%\label{Clebsch Lax matr}
L(\lambda)=\lambda A+M+\lambda^{-1}PP^{\rm T},\qquad
U(\lambda)=\lambda B+\Omega.
\end{equation*}

In \cite{S1} we found an integrable Lagrangian discretization of the flow of
the Clebsch system characterized by
\[
c_{jk}=\frac{1}{J_j+J_k},\qquad b_k=J_k,\quad {\rm so\ that}\quad a_k=J_k^2.
\]
This flow may be considered as a particular case of the motion of a rigid
body in a quadratic potential (notice that the kinetic energy term in
(\ref{Kirch Lagr}) is in this case typical for the heavy top, since
${\mathcal J}(\Omega)=J\Omega+\Omega J$). Here we concentrate on another flow of the
Clebsch system characterized by
\[
c_{jk}=1,\qquad {\rm so\ that}\quad a_k=b_k.
\]
The corresponding Lagrange function reads:
\begin{equation}\label{Clebsch spec Lagr}
{\mathbf L}(g,\dot{g})={\mathcal L}^{(l)}(\Omega,P)=
\frac{1}{2}\,\langle\Omega,\Omega\rangle+
\frac{1}{2}\,\langle P,BP\rangle,
\end{equation}
so that
\begin{equation*}%\label{Clebsch spec M}
M=\Omega,
\end{equation*}
and the equations of motion become
\begin{equation}\label{Clebsch spec}
\left\{\begin{array}{l} \dot{M}=P\wedge (BP),\vspace{1mm}\\
\dot{P}=-MP.\end{array}\right.
\end{equation}
The Hamilton function of this flow is given by
\begin{equation*}%\label{Clebsch spec Ham}
H(M,P)=\frac{1}{2}\langle M,M\rangle-\frac{1}{2}\,\langle P,BP\rangle.
\end{equation*}
The Lax representation of this flow can be given in two equivalent forms:
\begin{equation*}%\label{Clebsch spec Lax}
\dot{L}(\lambda)=[L(\lambda),U_+(\lambda)]=[U_-(\lambda),L(\lambda)],
\end{equation*}
where
\begin{equation*}%\label{Clebsch spec Lax matr}
L(\lambda)=\lambda B+M+\lambda^{-1}PP^{\rm T},\qquad
U_+(\lambda)=\lambda B+M,\qquad U_-(\lambda)=\lambda^{-1}PP^{\rm T}.
\end{equation*}

Consider the following discrete time Lagrange function approximating
(\ref{Clebsch spec Lagr}):
\begin{gather}
{{\mathbb L}}(g_k,g_{k+1})=\Lambda^{(l)}(P_k,W_k)\nonumber\\
\qquad = -\frac{1}{\epsilon}\,{\rm tr}\log
\left(2I+W_k+W_k^{-1}\right)+\frac{4}{\epsilon}\,
\log\left(1+\frac{\epsilon^2}{4}\langle P_k,BP_k\rangle\right).
\label{dClebsch spec Lagr}
\end{gather}
Here, as usual, $W_k=g_k^{-1}g_{k+1}$ and $P_k=g_k^{-1}p$.
\begin{theorem}
The reduced Euler--Lagrange equations of motion for the Lagrange function
(\ref{dClebsch spec Lagr}) read:
\begin{equation}\label{dClebsch spec}
\left\{\begin{array}{l}
M_{k+1}=M_k+\epsilon \displaystyle\frac{P_k\wedge(BP_k)}
{1+(\epsilon^2/4)\langle P_k,BP_k\rangle},\vspace{2mm}\\
P_{k+1}=\displaystyle\frac{I-(\epsilon/2) M_{k+1}}{I+(\epsilon/2)M_{k+1}}\,P_k.
\end{array}\right.
\end{equation}
The map $(M_k,P_k)\mapsto(M_{k+1},P_{k+1})$ is Poisson with respect to the
bracket (\ref{en LP br 1}), (\ref{en LP br 2}). This map is completely
integrable and admits the following Lax representation:
\begin{equation*}%\label{dClebsch spec Lax}
L_{k+1}(\lambda)=V_k(\lambda)L_k(\lambda)V_k^{-1}(\lambda),
\end{equation*}
with the matrices
\begin{equation*}%\label{dClebsch spec LV}
L_k(\lambda)=\left(I+\frac{\epsilon^2}{4}\,B\right)^{-1}\left(\lambda B+M_k+
\lambda^{-1}{\mathcal P}_k\right),\qquad
V_k(\lambda)=\frac{I+(\epsilon\lambda^{-1}/2){\mathcal Q}_k}
{I-(\epsilon\lambda^{-1}/2){\mathcal Q}_k},
\end{equation*}
where
\begin{gather}\label{dClebsch spec P}
{\mathcal P}_k=\left(I+\frac{\epsilon}{2}\,M_k\right)P_kP_k^{\rm T}
\left(I-\frac{\epsilon}{2}\,M_k\right),\\
%\label{dClebsch spec Q}
{\mathcal Q}_k=\frac{1}{1+\left(\epsilon^2/4\right)\langle P_k,BP_k\rangle}\,
P_kP_k^{\rm T}\left(I+\frac{\epsilon^2}{4}\,B\right).\nonumber
\end{gather}
\end{theorem}

\begin{proof} Derivation of equations of motion is based on
(\ref{dEL top left red}), (\ref{dEL top left M}). We have:
\[
M_{k+1}=\frac{2}{\epsilon}\cdot\frac{W_k-I}{W_k+I}\quad\Rightarrow\quad
W_k=\frac{I+(\epsilon/2)M_{k+1}}{I-(\epsilon/2)M_{k+1}},
\]
which proves the second equation of motion in (\ref{dClebsch spec}), and
\[
\nabla_P\Lambda^{(l)}(W_k,P_k)=\frac{\epsilon BP_k}
{1+\left(\epsilon^2/4\right)\langle P_k,BP_k\rangle},
\]
which proves the first one. As for the Lax representation, it is verified
by a direct calculation. It is most convenient to check it in the form
\begin{gather}
\left(I-\frac{\epsilon\lambda^{-1}}{2}{\mathcal R}_k\right)
\left(\lambda B+M_{k+1}+\lambda^{-1}{\mathcal P}_{k+1}\right)
\left(I+\frac{\epsilon\lambda^{-1}}{2}{\mathcal Q}_k\right)\nonumber\\
\qquad =\left(I+\frac{\epsilon\lambda^{-1}}{2}{\mathcal R}_k\right)
\left(\lambda B+M_k+\lambda^{-1}{\mathcal P}_k\right)
\left(I-\frac{\epsilon\lambda^{-1}}{2}{\mathcal Q}_k\right),\label{dClebsch spec aux1}
\end{gather}
where
\begin{equation*}%\label{dClebsch spec R}
{\mathcal R}_k=\frac{1}{1+(\epsilon^2/4)\langle P_k,BP_k\rangle}
\left(I+\frac{\epsilon^2}{4}\,B\right)P_kP_k^{\rm T}.
\end{equation*}
Let us indicate the main steps of this calculation. Expanding
(\ref{dClebsch spec aux1}) in powers of $\lambda$, we come to the following
equations:
\begin{gather}
\lambda^0:    \quad M_{k+1}-M_k=\epsilon({\mathcal R}_kB-B{\mathcal Q}_k);
\label{dClebsch spec aux3}\\
\lambda^{-1}: \quad {\mathcal P}_{k+1}-{\mathcal P}_k
=\frac{\epsilon}{2}\,{\mathcal R}_k(M_{k+1}+M_k)-
\frac{\epsilon}{2}\,(M_{k+1}+M_k){\mathcal Q}_k;
\label{dClebsch spec aux4}\\
\lambda^{-2}: \quad ({\mathcal P}_{k+1}+{\mathcal P}_k){\mathcal Q}_k-{\mathcal R}_k({\mathcal P}_{k+1}+{\mathcal P}_k)=
\frac{\epsilon}{2}\,{\mathcal R}_k(M_{k+1}-M_k){\mathcal Q}_k;
\label{dClebsch spec aux5}\\
\lambda^{-2}: \quad {\mathcal R}_k({\mathcal P}_{k+1}-{\mathcal P}_k){\mathcal Q}_k=0.
\label{dClebsch spec aux6}
\end{gather}
Here (\ref{dClebsch spec aux3}) follows from the first equation in
(\ref{dClebsch spec}). Equation (\ref{dClebsch spec aux4}) yields
(\ref{dClebsch spec aux6}), which follows from the skew-symmetry of
$M_{k+1}$, $M_k$. To prove (\ref{dClebsch spec aux4}) we notice
that the first equation of motion in
(\ref{dClebsch spec}) easily yields the following expressions:
\begin{equation}\label{dClebsch spec aux8}
{\mathcal Q}_k=P_kP_k^{\rm T}\left(I+\frac{\epsilon}{4}(M_{k+1}-M_k)\right),\qquad
{\mathcal R}_k=\left(I-\frac{\epsilon}{4}(M_{k+1}-M_k)\right)P_kP_k^{\rm T},
\end{equation}
while the second equation of motion in (\ref{dClebsch spec}) shows that
\begin{equation}\label{dClebsch spec aux9}
{\mathcal P}_{k+1}=\left(I-\frac{\epsilon}{2}\,M_{k+1}\right)P_kP_k^{\rm T}
\left(I+\frac{\epsilon}{2}\,M_{k+1}\right).
\end{equation}
This is a companion formula to (\ref{dClebsch spec P}).
Now (\ref{dClebsch spec aux4}) follows directly from (\ref{dClebsch spec aux8}),
(\ref{dClebsch spec aux9}), (\ref{dClebsch spec P}). Another consequence
of these three formulas is:
\begin{gather*}
({\mathcal P}_{k+1}+{\mathcal P}_k){\mathcal Q}_k=
{\mathcal R}_k({\mathcal P}_{k+1}+{\mathcal P}_k)\\
\qquad =
\left(I-\frac{\epsilon}{4}(M_{k+1}-M_k)\right)P_kP_k^{\rm T}
\left(I+\frac{\epsilon}{4}(M_{k+1}-M_k)\right),
\end{gather*}
and this proves (\ref{dClebsch spec aux5}), since the right-hand side
of this formula vanishes due to the skew-symmetry of $M_{k+1}$, $M_k$.
\end{proof}

We conclude this section with the following remarks. The discrete time system
introduced here was found in the case $n=3$ and in the ${\rm su}(2)$ framework
by Adler \cite{Ad}, who however did not use the Lagrangian formalism. It is
well known that the restriction of the flow (\ref{Clebsch spec}) to the
symplectic leaf in ${\rm e}^*(n)$ of the smallest possible dimension $2n$ is
equivalent to the famous {\it Neumann system} which describes the motion of a
particle on the surface of the sphere $S=\{x\in\mathbb R^n: \langle x,x\rangle
=1\}$ under the influence of the harmonic potential $(1/2)\langle x,Bx\rangle$.
(The identification of the above mentioned symplectic leaf with $T^*S$ is
achieved via the formulas $P=x$, $M=x\wedge p$). It turns out that on this
symplectic leaf our map (\ref{dClebsch spec}) leads to {\it Adler's
discretization of the Neumann system}~\cite{Ad}, whose complete integrability
was proved in~\cite{S2}. The present construction delivers therefore a Lax
representation for this discretization of the Neumann system -- the problem
left open in~\cite{S2}.

\section{Conclusion}
The models introduced in the present paper serve as further important
examples of completely integrable Lagrangian systems with a discrete time
\`a la Moser--Veselov. The version of the discrete Lagrangian reduction
leading to systems on duals to semidirect product Lie algebras was
developed in \cite{BS2}. The list of relevant examples grows slowly,
but now it already became quite representative: \cite{V,MV,BS1,S1},
and the present work. Several remarks are in order here.
\begin{itemize}
\itemsep0mm
\item Usually, discrete time Lagrangians lead to multi-valued maps
(correspondences), cf.~\cite{MV}. This is also the case for the models
introduced in \cite{V,S1} (discrete time Euler top and a
discretization of the rigid body dynamics in a quadratic potential).
Amasingly, the models of \cite{BS1}, as well as those of the present paper,
lead to {\it single-valued}, and moreover {\it explicit} maps.
Of course, one would like to be able to predict this outstanding property
by just looking at the Lagrange function, but at present we do not know how
to do this, since no deep reasons for this behavior are apparent.
\item A very intriguing and still not completely
understood point is the capability of the discrete Lagrangian approach to
produce completely integrable systems. It should be stressed that all the
models listed above were found by guess, and the same holds for their Lax
representations. There seems to exist (at least, at present) no regular
procedure for finding decent integrable discretizations for finite-dimensional
systems of the classical mechanics, like the rigid body dynamics.
This is in a sharp contrast to the area of integrable
{\it differential-difference}, or {\it lattice} systems,
where the problem of integrable discretization may be solved in an
almost algorithmic way, preserving the Lax matrix and the Hamiltonian
properties, cf.~\cite{S0}.
\item For any completely integrable system, the involutive integrals of motion
yield a set of commuting Hamiltonian flows (called a {\it hierarchy} in the
infinite dimensional situation). These flows share the Lax matrix with the
original one, and there exist well-established procedures for finding the
Lax representations for them. In the discrete time, the situation seems to be
different. Although there is a recipe for producing commuting Poisson maps
for a given one, provided its Lax representation admits an $r$-matrix
interpretation (which is the case for all the models listed above), there is
no way, in general, for producing decent discretizations for concrete
flows. Here ``decent'' means given by nice formulas in physical coordinates,
or, if one prefers more precise terms, given by explicit Lagrange functions
on Lie groups. In this connection it should be mentioned that in the only
case when decent discretizations are known for two different flows of the same
hierarchy (the Clebsch problem, cf.~\cite{S1} and the present paper), these
discretizations belong to {\it different} hierarchies and posess {\it different}
Lax matrices and integrals of motion.
\end{itemize}

All in one, it should be said that while the Veselov's papers \cite{V,MV}
appeared more than a decade ago and happened to give a strong
impetus for the development of the whole subject of discrete integrable systems,
the narrower area of integrable discrete Lagrangian systems pioneered there
is still at the beginning of its development.

Generally, we consider the discrete time Lagrangian mechanics as an important
source of symplectic and, more general, Poisson maps. From some points of view
the variational (Lagrangian) structure is even more fundamental and important
than the Poisson (Hamiltonian) one. (Cf. \cite{HMR,MPS}, where a similar
viewpoint is represented. Notice that the book~\cite{K} also dealing with the
omnipresence of Hamiltonian systems on the semidirect product Lie algebras,
takes an opposite viewpoint: the variational (Lagrangian) principles are
derived there from the postulated Hamiltonian structures.) It would be
important to continue the search for integrable Lagrangian discretizations of
the known integrable systems. Also generalizations to the infinite dimensional
case, e.g. to a discretization of ideal compressible fluids motion
(see \cite{HMR,K}), are highly desirable.

\subsection*{Acknowledgements}
I would like to express my gratitude to A~Bobenko, the collaboration with
whom in \cite{BS1,BS2} initiated this work.

Thanks are due also to B~Kupershmidt for a number of useful remarks.

The results of this paper were the subject of my invited talk at the
SIDE IV conference (Tokyo, 2000). I cordially thank J~Satsuma and the
University of Tokyo for their hospitality during this exciting meeting.

This research was financially supported by the DFG (Deutsche
Forschungsgemeinschaft)
in the frame of SFB 288 ``Differential Geometry and Quantum Physics''.

\label{suris-lastpage}

\end{document}